\documentstyle[epsf,prl,multicol,aps]{revtex}

\newcommand{\ccmo}{CaCu$_3$Mn$_4$O$_{12}$}

\begin{document}
\title{Magneto-electronic Properties of a Ferrimagnetic Semiconductor: \\
The Hybrid Cupromanganite CaCu$_3$Mn$_4$O$_{12}$}
\author{Ruben Weht$^{\dag}$ and Warren E. Pickett$^{\ddag}$}
\address{$^{\dag}$Departamento de F\'{\i}sica, CNEA,
Avda. General Paz y Constituyentes, 1650 - San Mart\'{\i}n,
Argentina}
\address{$^{\ddag}$Department of Physics, University of 
California, Davis CA 95616}
\date{\today}
\maketitle

\begin{abstract}
The mixed manganite-cuprate CaCu$_3$Mn$_4$O$_{12}$ ~is found, using density
functional methods, to be a narrow gap (90 meV calculated) 
ferrimagnetic semiconductor.
Cu (formally S=$\frac{1}{2}$) antialigns with Mn (formally S = $\frac{3}{2}$), 
and the net spin moment is 9 $\mu_B$ consistent with the formal spins.
Holes have Cu $d_{xy}$-O $p_{\sigma}$ 
({\it i.e.} antibonding $dp\sigma$) character
with spins aligned antiparallel to the net magnetization; electrons have the
opposite spin and have mixed Cu $d_{xy}$ -Mn $e_g$ 
character.  Thermally excited electrons and holes will each be fully
spin polarized, but in opposite directions.  The properties of this material
are strongly tied to the distorted quadruple
perovskite structure, which is closely
related to the skutterudite structure. The observed resistivity,
magnetoresistance, and magnetization are discussed in terms of our results.
\end{abstract}

\vskip 1cm 

\begin{multicols}{2}

\section{Introduction}

Reports in the past few years of very high values of magnetoresistance
(MR)  in manganite perovskites have stimulated a tremendous interest in
those compounds~\cite{1993}, already well known from seminal works in the
1950s. These colossal magnetoresistance (CMR) materials are
attracting a level of attention only rivaled by the continued interest in
the high temperature cuprates, which are based on a closely related transition
metal oxide perovskite structure. With the development of more and more complex
materials by both conventional growth techniques and artificial,
non-equilibrium growth procedures, it was inevitable that the combination
of Cu and Mn oxides would be explored.

From the basic point of view, attractiveness of these systems is due to 
the richness of their very complex phase diagrams, which in both cases may
be viewed from the undoped antiferromagnetic (AFM) insulator (I) parent
compound. In manganites, {\it viz.} La$_{1-x}$Ca$_{x}$MnO$_{3}$, the phase
diagram shows a strong relation between lattice, spin, charge and orbital
degrees of freedom. From the La-rich end, the AFM-I phase gives way to a
ferromagnetic (FM) metal, while from the Ca-rich side the system
remains insulating down to $x=0.50$ while displaying a variety of charge,
spin, and orbitally ordered phases. In layered cuprates the AFM-I phase
gives way to a metallic phase that becomes superconducting at remarkably
high temperatures, over 100 K in several cases. 

The high values of MR observed in the manganites may become important as
components of magnetoelectronic devices. However, before widespread
practical applications can be made some of their properties have to be
improved. In particular, there are two limiting factors in their
performance: one is that the highest values of MR can be reached only at
relatively low temperatures and high magnetic fields (typically 250 K and
several Tesla for La$_{3/4}$Ca$_{1/4}$MnO$_3$), the second is that
the magnetoresistance is large only in a narrow interval around the
Curie point.

Both from the point of view of the manganite-cuprate combination and for
the possibility of improving device characteristics, the recent report by
Zeng {\it et al.}\cite{zeng} constitutes an interesting new development.
This group has shown that the hybrid cupromanganite
CaCu$_3$Mn$_4$O$_{12}$ (CCMO), which is an ordered perovskite with
quadrupled primitive unit cell, presents a quite large MR in 
polycrystalline samples (up to several
tens of percent) at low magnetic fields without the presence of either
mixed valency or metal-insulator transition. Moreover, this colossal MR
extends over a wide temperature range.

This compound contains two magnetic ions, formally given as Mn$^{4+}$
($d^{3},S=\frac{3}{2}$) and Cu$^{2+}$ ($d^{9},S=\frac{1}{2}$), on two
crystallographically distinct sites. Both its resistivity $\rho (T)$ and
magnetization $M(T)$, measured on polycrystalline samples, show
unconventional temperature dependencies. $\rho (T)$ is semiconducting in
behavior, but has {\it no} visible anomaly at the
Curie temperature T$_{C}$ = 355 K. $M(T)$, on the other hand, shows a
steep, nearly first-order-like jump at T$_{C}$ to $\sim $80\% of its
saturation value. The temperature dependence of the activated resistivity
(over a limited temperature range) suggests an energy gap of $\sim $0.12
eV.

The basic electronic, and even magnetic, structures of this compound are
not known. In this work we present the electronic features of this
material, relating them to its transport and magnetic properties. In Sec.
II we describe the distorted quadruple perovskite structure and outline
our method of calculation. The results for the band structure, density of
states, and magnetism are given in Sec. III. In Sec. IV we discuss the
avenues for magnetic coupling in CCMO, and magnetotransport is discussed
briefly in Sec. V. Our results are summarized in Sec.
VI.

\section{Crystal Structure and Method of Calculation}

\subsection{Structure}

The crystal structure, shown in Fig.~1 and discussed in detail by Chenavas
{\it et al.} 25 years ago~\cite{chenavas}, is a strongly distorted ordered
version of a perovskite with formula Ca$_{1/4}$Cu$_{3/4}$MnO$_{3}$. Formal
valence ideas suggest a Cu$^{2+}$ ion, so the system is isovalent with
CaMnO$_{3}$ and hence expected to be insulating, as observed.  Here the
Cu$^{2+}$, but not the Mn$^{4+}$, ion is expected to be a Jahn-Teller ion,
leading to an oxygen sublattice that corresponds to a tilted
three-dimensional network of MnO$_{6}$ octahedra. The Mn-O-Mn angle
becomes $\approx 142^{o},$ instead of 180$^{o}$ as in the ideal perovskite
structure. Two types of polyhedra are present at the A position: a 
slightly distorted O icosahedron around the Ca site and a roughly square
planar O-coordinated Cu site with a Cu-O distance of 1.94 ${\rm {\AA }}$.
The quadrupled perovskite cell has space group $Im3$ ($\equiv
I2/m\bar{3}$, No. 204 in the International Tables).

The quadrupling and distortion of the perovskite structure leads to low
site symmetries. The cations nevertheless all sit on the usual ideal
positions of the cubic perovskite lattice. The symmetry lowering arises,
first, from the replacement of 75\% of the Ca (in CaMnO$_{3}$) by Cu and 
second, by the
rotation of the O octrahedra which leaves O sites of the form
$(0,y,z)$$(y=0.3033,z=0.1822)$ and mirror site symmetry only. The Mn site at
$(\frac{1}{4},\frac{1}{4},\frac{1}{4})a$ has $\bar{3}$ symmetry, because
each MnO$_{6}$ octahedron rotates to orient a triangular face
perpendicular to a $<111>$ direction.  The Cu site retains $mmm$
symmetry, while the Ca site is in a $m3$ position.  The lattice constant
$a$ is approximately twice the related simple perovskite cell, and the
Bravais lattice is bcc.

This structure is closed related to that of skutterudite
CoSb$_{3}$\cite{skut}. Both are members of a class that can be generally
denoted as A$^{\prime }$A$_{3}^{\prime \prime }$T$_{4}$D$_{12}$, where
A$^{\prime }$, A$^{\prime \prime }$ are cations, T is a metal (usually
transition metal) ion, D is an anion, and the underlying structure is
perovskite.  
The relation of CoSb$_3$ to the CaCu$_{3}$Mn$_{4}$O$_{12}$ structure is:  
Co$\rightarrow $Mn; Sb$\rightarrow $O; the Sb$_{4}$ square becomes the
CuO$_{4}$ square; the Ca ion fills the La site in filled skuterudites (like
LaCo$_{4}$Sb$_{12}$).  The Cu site is not filled in the skutterudites.

It must be noted for the analysis that follows, that the CuO$_{4}$
``square'' is in fact not actually a square although all Cu-O bondlengths
are equal and all O sites are equivalent. 
Rather it is a rectangle whose O-Cu-O angles are 85.6$^{o}$ and
94.4$^{o}$. As a result the Cu d$_{xy}$, d$_{yz}$, and d$_{zx}$ orbitals
are not related by symmetry.
Nevertheless, we will use the term ``CuO$_{4}$ square'' and
when we discuss it, we will have in mind a local coordinate system in
which it lies in the $x-y$ plane.  

\begin{figure}[tbp]
\epsfxsize=8.5cm\centerline{\epsffile{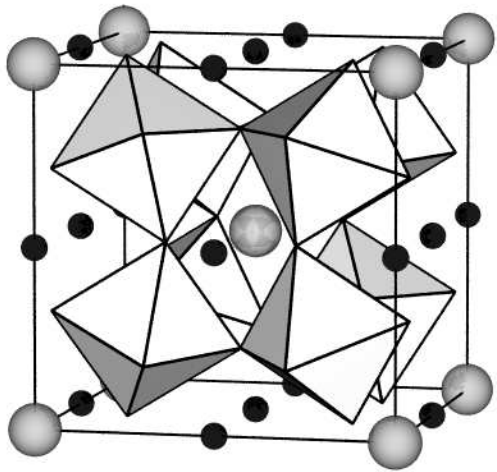}}
\caption{The crystal structure of the quadruple perovskite compound
CaCu$_3$Mn$_4$O$_{12}$.
Two primitive cells of the bcc structure are shown. Oxygen
atoms lie at the vertices of the MnO$_6$ octahedra.  Small black spheres
denote Cu atoms, large gray spheres are Ca atoms.}
\label{Fig1}
\end{figure}

Due to the low site symmetry arising from the distortion, the Mn
$d$-orbitals cannot be exactly associated with the standard
t$_{2g}$/e$_{g}$ symmetries. Instead, the $\bar 3$ symmetry of Mn dictates
that, in the local coordinate system where the threefold axis is the $\hat
z^{\prime}$ axis, the irreducible representations of the five $3d$ states 
are $z^{\prime 2},
\{x^{\prime}z^{\prime},y^{\prime}z^{\prime}\}$, and 
$\{x^{\prime}y^{\prime},x^{\prime 2}-y^{\prime 2}\}$.  
We will call in this work the standard
decomposition as ``local frame" and the last, symmetry-deterimined, 
one the ``symmetry adapted
frame".

\subsection{Calculations}

We have applied the linearized augmented plane wave method\cite{djsbook}
(LAPW)
that utilizes a fully general shape for density and potential. The WIEN97
code \cite{wien} has been used in the calculations. The experimental
lattice constant of 7.241~\AA = 2 $\times$ 3.62~\AA ~was used. 
LAPW sphere radii (R) of 1.90
a.u. were chosen for the Cu and Mn atoms, 2.00 a.u. for Ca and 1.60 a.u
for O, with cutoffs of RK$_{max}$ up to 7.0, providing basis sets with
more than 1800 functions per primitive cell. Self-consistency was carried
out on k-points meshes of up to 60 points in the irreducible Brillouin
zone (1000 points in the complete BZ). The generalized gradient
approximation (GGA) exchange-correlation
functional of Perdew {\it et al.} \cite{gga} was used in the present work,
except where noted.

A previous study of the electronic structure of CCMO has been reported by
Wu, Zheng, and Gong~\cite{wu} using both LDA and the LDA+U (corrections due
to strong on-site repulsion) 
within a local combination of atomic orbitals scheme.  They concluded that
LDA gives a metallic behavior, in disagreementt with the experimental
situation, and that it was necessary to apply the LDA+U method
to open a band gap.  We will show
here that the semiconducting behavior is well reproduceed just considering
GGA within density functional theory. 
Discrepancies between our results and those of Wu {\it et al.} are due to
approximations made in their methods, and possibly to their choice of 
basis set (see 
\cite{comment}).

\section{Computational Results}

\subsection{Band Structure}
The important valence-conduction band region contains over 60 bands, hence
the full band structure will not be presented.
The majority and minority band structures bounding the gap are shown in Fig.
2, and the densities of states (DOS) in Fig. 3. The
calculated gaps are 0.50 eV for spin up (majority) and of 0.18 eV for spin
down (minority), and each is direct. 
For the minority carriers the gap occurs at H
between pure Cu $d_{xy}$ states below the gap to pure Mn $t_{2g}$ character
above. The thermal gap is quite small, 0.09 eV, and is both
{\it indirect} and spin-asymetric ({\it i.e.} bounded by
the spin down valence band maximum at H and the spin up conduction band
minimum at $\Gamma$). It is common for the band gap in density
functional calculations to be
smaller than the true gap, but the band character and shape on either side
of the gap are nonetheless given reasonably. In this case the calculated gap
is quite similar to the experimental estimate.

The bands at the valence band maximum (spin down) have a strong Cu $d$
character. They arise from the $dp\sigma$ antibonding
interaction of the d$_{xy}$ (see below) with O p$_\sigma$ orbitals in the 
CuO$_4$ square, and are the analog of the d$_{x^2-y^2}$ states in the layered
cuprates. This bonding makes a band with a dispersion of almost 1.5 eV. The
other Cu $d$ orbitals are concentrated in a small region between 1 and
2.5 eV below the gap.
As expected for the Mn$^{4+}$ $(d^3)$ ion the corresponding ${\rm t_{2g}}$
and ${\rm e_g}$ orbitals (in the local octahedron-adapted coordinates) are 
well separated, with the ${\rm t_{2g}}$ majority orbitals filled.

The dispersion of the bands bounding the gap influences transport
coefficients. 
As discussed above, the valence bands are derived primarily
from a $dp\sigma$ antibonding combination of the Cu d$_{xy}$ and the four
neighboring O $p\sigma$ orbitals, which we will denote ${\cal D}_{xy}$. 
If the CuO$_4$ unit were actually a square, ${\cal D}_{xy}$ would not couple
with other Cu $d$ -- O $p$ combinations. Since the unit is not exactly a
square, the Cu $d_{xy}$ orbital will also mix with inplane O $p_{\pi}$
orbitals via $dp\pi$ coupling, which may not be negligible.
From the depiction of the CuO$_4$ squares in Fig. 4, it can be seen that the
${\cal D}_{xy}$ functions on
neighboring squares are orthogonal by symmetry. As a result, the dispersion,
which is 1.2 eV along H-$\Gamma$, cannot develop from direct ${\cal D}_{xy}$
-- ${\cal D}_{xy}$ hopping.

\begin{figure}[tbp]
\epsfxsize=8.5cm\centerline{\epsffile{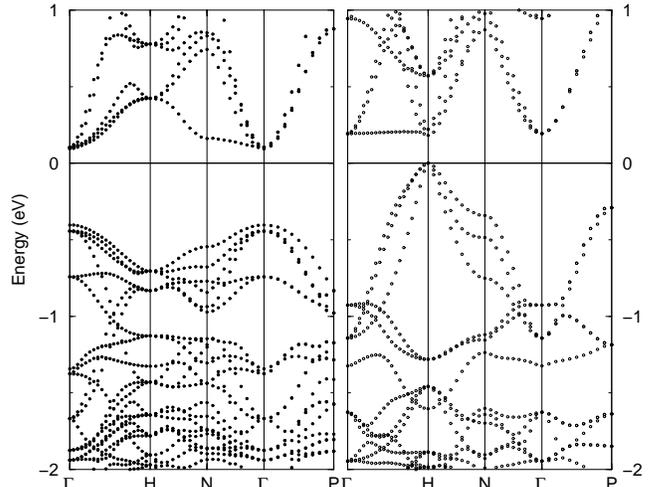}}
\caption{Band structures for spins up (left frame) and down (right frame). 
The symmetry points are $\Gamma$ = (0, 0, 0), H = 
$\frac{2\protect\pi}{a}$ (1, 0, 0), N = $\frac{2\protect\pi}{a}(\frac{1}{2},
\frac{1}{2}$, 0), P = $\frac{2\protect\pi}{a}(\frac{1}{2},\frac{1}{2},
\frac{1}{2})$.}
\label{Fig2}
\end{figure}

\begin{figure}[tbp]
\epsfxsize=8.0cm\centerline{\epsffile{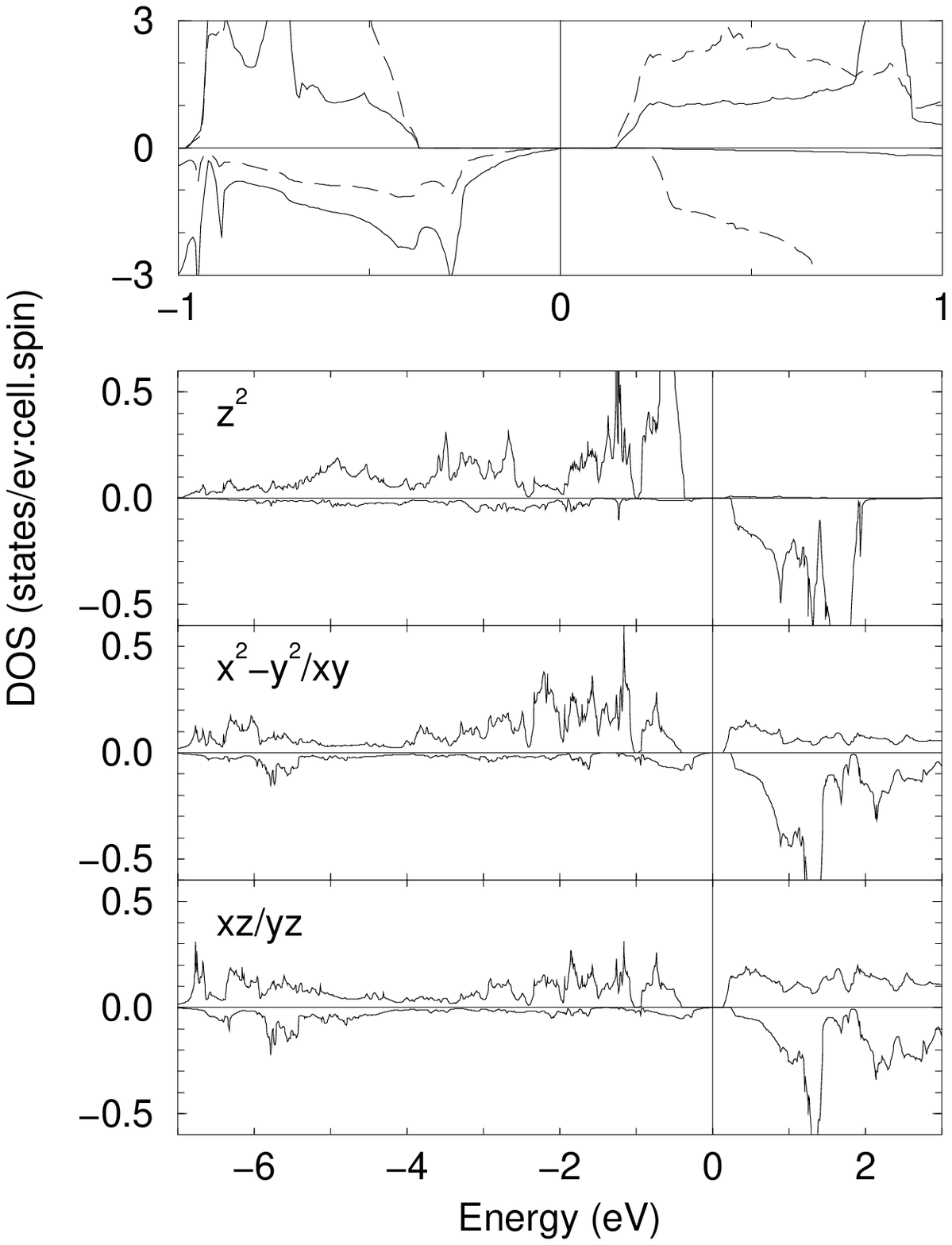}}
\caption{Bottom: Partial density of states (DOS) for the Mn atom, decomposed
according the symmetry adapted frame, which in the text is denoted with
primes ($x^{\prime}y^{\prime}$, etc.). Top: A detail of the 
partial density of states  around the gap, with Cu as solid line and Mn as 
dashed line.  The majority DOS is plotted upward, the minority is plotted
downward.
The zero of energy lies within the gap. }
\label{Fig3}
\end{figure}

The symmetry character of the Mn $d$ shell is important to identify, since
it affects magnetic coupling. The natural presumption is that, since the 
MnO$_6$ octahedron is not strongly distorted, then with respect to the local 
coordinate frame in which the axes extend from Mn through the neighboring O
ions (or nearly so), the configuration can be described as 
$t_{2g\uparrow}^3$, the fully polarized $d^3$ configuration. 
While this picture is certainly
roughly true, there is a related and precise aspect to the exact site
symmetry of Mn. 
The Mn $d$ DOS, decomposed into the symmetry adapted frame (Sec. II.A)
indicates that it is the $z^{\prime 2}$ state alone that is completely
occupied in the majority bands, and very nearly completely unoccupied in the
minority bands. The other two twofold irreducible representations are
strongly mixed in both the majority occupied and unoccupied states. 
This full polarization of the $z^{\prime 2}$ orbital reflects the fact
that it is directed along a $<111>$ axis and therefore is an equal 
combination of (only) the $t_{2g}$ orbitals in the 
local frame: $z^{\prime 2} = (xy+yz+zx)/\sqrt 3$.
Thus the $t_{2g}$ orbitals are fully occupied as expected.

\subsection{Magnetic Structure}

The calculations were initiated with magnetic Cu and Mn ions, aligned in
ferromagnetic fashion. Both Mn and Cu ions remain magnetic, but the Cu
moments strongly favor being {\it antiparallel} to the Mn moments, resulting
in a ferrimagnetic spin ordering. Both the Mn moments ($S=\frac{3}{2}$,
ideally 3 $\mu_B$) and the Cu moment ($S=\frac{1}{2}$, ideally 1 $\mu_B$)
are reduced by hybridization with the O 2$p$ states, with resulting
moments being near 2.42 $\mu_B$ and -0.45 $\mu_B$ respectively. The net
moment is 9 $\mu_B$ per formula unit, which is what would be obtained
from the formal moments aligned ferrimagnetically (4 Mn $\times~3\mu_B$
-~3 Cu $\times~1\mu_B$). 
The magnetic moment on the Cu ion is a result of the exchange splitting on
the ${\cal D}_{xy}$ orbital, which in this coordinate system is the orbital
with lobes pointing to the nearest neighbor oxygens. The almost
electronically isolated CuO square favors this configuration as was shown
previously in other zero or one dimensional copper oxides \cite
{li2cuo2,la2bacuo5}. Magnetism in the Mn ion comes as expected from the
splitting of the t$_{2g}$ orbitals and filling only of the spin up states.

Experimental study of the magnetic arrangement in CaCu$_3$Mn$_4$O$_{12}$~
has been reported.
Neutron scattering has established that some
samples are non-stoichiometric, with the (also Jahn-Teller active) Mn$^{3+}$
ion substituting partially on the Cu site.\cite{doped} It was
suggested that in these samples the
order may be ferromagnetic or canted. In a nearly stoichiometric sample, the
total magnetization decreased and ferrimagnetic ordering was inferred.

To investigate how strong this ferrimagnetic configuration is, we have
done fixed spin moment calculations that impose ferromagnetic alignment
of the spins. Due to the Mn$^{4+}$ assignment the
magnetic moment of the idealized Mn ion in this compound is 3 $\mu_B$, so 
we fixed the total moment to 15 $\mu_B$ to include aligned moments of
the three Cu ions. 
Our results show that this state
is almost 1 eV higher in energy that the ground state, with
little change of the individual magnetic moments. Supposing a near neighbor
spin Hamiltonian with a Cu-Mn coupling of the form 
\begin{equation}
H^{Cu-Mn} = \sum_{<ij>} J^{Cu-Mn}_{ij} \hat S^{Cu}_i \cdot \hat S^{Mn}_j,
\end{equation}
with the sum being over pairs, the three Cu ions per cell and eight Mn
neighbors per Cu would lead to an AFM exchange coupling of $J^{Cu-Mn} = $20
meV. 

\begin{figure}[tbp]
\epsfxsize=8.0cm\centerline{\epsffile{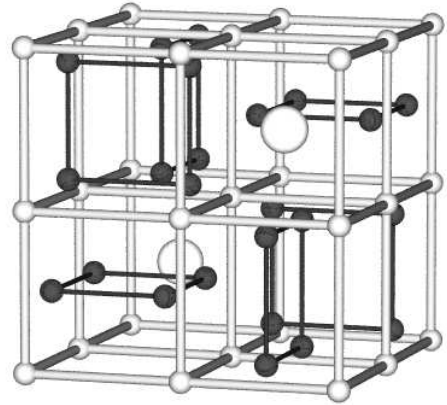}} 
\caption{A skeleton drawing of the CaCu$_3$Mn$_4$O$_{12}$ ~structure,
emphasizing the CuO$_4$ `squares.' Small light grey dots represent the Mn ions,
while the squares represent the CuO$_4$ units with black O ion. The Cu ion
lies at the center of the square and is not shown. The large light gray
sphere denotes Ca. This figure is displaced by $(\frac{a}{4},\frac{a}{4}, 
\frac{a}{4}$) from the regions shown in Fig. 1. }
\label{Fig4}
\end{figure}

The magnetism of CaCu$_3$Mn$_4$O$_{12}$ presents several interesting
questions. First of all, the exchange splitting of the Cu d$_{xy}$ state is $%
\Delta^{Cu}_{ex} \sim$ 1.2 eV from the Cu $d_{xy}$ DOS (not shown). 
Given its moment of $m_{Cu}$%
=0.45 $\mu_B$, this gives a ratio $\Delta^{Cu}_{xy}$/$m_{Cu}$ = 2.7 eV/$\mu_B
$, a very large value considering this ratio (roughly the Stoner $I^{Cu}_{xy}
$) is usually less than unity in transition metal magnets. The exchange
splitting of the other Cu d orbitals is smaller, of the order of 0.4 eV.
This difference reflects a strongly anisotropic exchange potential on the Cu
ion.

The exchange splitting on the Mn ion is harder to estimate, due to the
strong dissimilarity of the majority and minority $d$ DOS (Fig. 3a). This is a
rather common occurrence in ferr{\it i}magnets, where the spin up bands
cannot be regarded as an exchange shifted version of the spin down bands,
even to zeroeth order.

\section{Magnetic Coupling}

The magnetic coupling in CaCu$_3$Mn$_4$O$_{12}$~ is potentially quite
complex to unravel due to the two types of magnetic ions and the large cell
with low symmetry sites. The bond angles involved in the various exchange
processes in CaCu$_3$Mn$_4$O$_{12}$~ are: O-Cu-O, 85.6$^{\circ}$ and 94.4$%
^{\circ}$; O-Mn-O, 90.0$^{\circ}$; Mn-O-Mn, 142$^{\circ}$; Cu-O-Mn, 108.7$%
^{\circ}$. Nearest neighbor bond distances are $d$(Mn-O) = 1.915 \AA, $d$%
(Cu-O)=1.942 \AA, $d$(Ca-O)=2.56 \AA. Note that in the undistorted structure
with twelvefold coordinated Cu, $d$(Cu-O) would be $\sqrt 2a/4$ = 2.56 \AA; {%
\i.e.} the rotation of the MnO$_6$ octahedra decreases the Cu-O distance by
0.62 \AA.

Lacroix\cite{lacroix} has discussed the couplings in the isostructural
compound CaCu$_3$Ti$_4$O$_{12}$, where Ti is $d^0$ and there are only Cu-Cu
interactions to consider. The observed magnetic 
order is antiferromagnetic,\cite{canted}
but with a three sublattice, non-collinear structure in which three Cu spins
in a (111) plane have projections on that plane that lie at 120$^{\circ}$
angles to each other. The three Cu spins in the next (111) layer are
antiparallel to the first three. Lacroix argues that, because the exchange
coupling path is through the TiO$_6$ octahedron, first, second, and third
neighbor Cu-Cu coupling $J_m^{Cu-Cu}$ ($m$=1,2,3) may be comparable in
magnitude. This should be true if direct O-O hopping within a MnO$_6$ 
octahedron or within a CuO$_4$ square is neglected.
Non-collinearity of the spins was attributed to spin-orbit coupling. In CaCu$%
_3$Mn$_4$O$_{12}$~ the Cu-Cu geometry is the same, so all three couplings
should again be considered. However, the MnO$_6$ octahedron will provide
different coupling than would the TiO$_6$ tetrahedron.

In addition, Mn-Mn coupling is essential as well as the Mn-Cu coupling
discussed in the previous section. Since Mn spins lie on a simple cubic
lattice connected by a single O$^{2-}$ ion, the Goodenough-Kanamori-Anderson
(GKA) rules can be applied to understand $J^{Mn-Mn}$. Further Mn-Mn
couplings should be small and not affect qualitative behavior. Whereas 180$%
^{\circ}$ Mn$^{4+}$-O-Mn$^{4+}$ coupling is antiferromagnetic ({\it viz.}
CaMnO$_3$), when this angle is reduced to 142$^{\circ}$ a ferromagnetic sign
is expected. Parallel alignment of the Mn spins is observed, and is the only
situation we have considered in our calculations.

Finally, there is the question of Mn-Cu coupling. In this structure all
oxygen ions are equivalent and are coordinated with two Mn ions and one Cu
ion. We assume that only a single Mn-Cu coupling is important; this interaction
connects a Cu spin to eight Mn nieghbors and each Mn spin to six neighboring
Cu spins. The ferromagnetic/ferrimagnetic difference in energy in the last
section identified $J^{Mn-Cu}$ = 20 meV (antiferromagnetic).

\section{Discussion}

The picture we obtain leads to a ferrimagnetic semiconductor with a calculated
indirect gap of 90 meV. Electron carriers and hole carriers will have
oppositely directed spins (up and down, respectively, with respect to the
net macroscopic magnetization $\vec M$). Here we consider the experimental
data of Zeng {\it et al.} in the light of our predicted electronic and
magnetic structure.

{\it Resistivity and MR.} Zeng {\it et al.} interpreted their data in the
neighborhood of T$_C$ in terms of a gap of around 120 meV. However, one
should note that their measured resistivity does not behave in activated
fashion over any extended temperature range; 
rather, in the range 25 -- 300 K the resistivity of their sample
can be described by the form 
\begin{equation}
\rho(T) = \rho_o e^{-T/T_o};~~\rho_o = 10^5 \Omega cm, ~~T_o = 180 K.
\end{equation}
Below 25 K $\rho$ deviates only slightly (upward) from this form. Since this
data is from polycrystalline samples, $\rho(T)$ may be affected strongly by
extrinsic factors, and the data would give only an upper bound on the
intrinsic resistivity.  Moreover, the variation of M(T) implies variation
in carrier density with temperature due to the dependence of the gap on
the ordered moment, complicating the interpretation of $\rho(T)$.

In a stoichiometric single magnetic domain sample of CaCu$_3$Mn$_4$O$_{12}$~
in zero field, the equal number of thermally excited 
electron and hole carriers fixes the
position of the chemical potential within the gap.  In this compound,
each type of carrier is fully polarized (electrons, up; holes, down) and
there will be no spin
scattering due to the absence of available spin-flipped states for each type
of carrier. An applied field produces a relative shift of the up and down
bands by $g \mu_B B$, where $g$ is the average of the electron 
and hole g factors. If $\vec B$ is applied parallel to $\vec M$,
the gap is narrowed; however, due to the requirement of equal number of
electrons and holes there will be no induced magnetization (and therefore no
spin susceptibility as long as $k_B T$ is much smaller than the gap). 
The field-induced change in the carrier density $n$ follows the form 
\begin{equation}
n(B,T) = n(B=0,T)~e^{\frac{g\mu_B B}{k_B T}}.
\end{equation}

Even if the measured $\rho(T)$ is not intrinsic, as long as it is
proportional to the number of excited carriers, at low T part the
field-induced change in carrier density will contriubte to the negative MR. 
If $g \sim$ 2, the enhancement in the carrier density
for B = 6 T (the highest field measured) is 35\% for
T=25 K and 2.6\% for T = 300 K. 
The (negative) MR
at 5 T was reported to be 35\% and 6-7\% respectively. Since at B = 5 T the
magnetization will surely be aligned with the field, the predicted decrease
in gap and increase in carrier density with field can account for much of
the observed MR. The temperature dependence of $\rho$
remains to be understood.

{\it Magnetization M(T).} Zeng {\it et al.} reported a normal hysterisis
loop characteristic of a soft magnet, but the magnetization $M(T)$ shows an
almost first-order-like jump to $\sim 75\%$ of the saturation magnetization
just below T$_C$, and no observable structure in $\rho(T)$ at T$_C$. This
behavior of M(T) and $\rho(T)$
is likely to be strongly affected by the granularity of the
sample.

\section{Summary}
Accurate band structure calculations predict that \ccmo~is a very small
indirect gap ferrimagnet, with Cu spins antialigned with the Mn spins and
a net moment of 9 $\mu_B$ per formula unit.  The gap lies between majority
spin conduction states and minority spin valence bands.  As a result,
application of a magnetic field parallel to the magnetization decreases
the gap, and this effect will contribute to the low temperature 
magnetoresistance.  This unusual spin arrangement around the gap makes
it attractive for possible applications in spin electronics devices.

The basic electronic structure conforms to expectations based on the
formal valence: the Cu$^{2+}$ ion is polarized due the $dp\sigma$
antibonding interaction with the surrounding four oxygen ions, and the 
Mn$^{4+}$ ion has fully occupied and fully polarized $t_{2g}$
orbitals, and no Jahn-Teller instability.  The fourfold coordinated
Cu ion distorts the perovskite structure, but in a way that constrains
the normal floppy nature of the perovskite crystal structure.

An antiferromagnetic coupling $J^{Cu-Mn} \approx$ 20 meV between the 
Cu and Mn spins is obtained from a fixed-spin-moment calculation.  The
ferromagnetic coupling between neighboring Mn ions can be rationalized
by the strong departure of the Mn-O-Mn angle from 180$^{\circ}$ (it is
142$^{\circ}$) but a quantitative understanding of the magnetic coupling
is still lacking.

\section{Acknowledgments}

We thank the hospitality of the Institute for Theoretical Physics,
University of California at Santa Barbara, where part of this work was done
(supported by National Science Foundation Grant PHY94-07194). We also
acknowledge discussions with B. L. Gyorffy, N. A. Hill, D. Johnston, D.
Khomskii, and F. Parisi. 
W.E.P. was supported by the National Science Foundation under
Grant DMR-9802076 and by Office of Naval Research Grant No.
N00014-97-1-0956. Computational work was supported by the National
Partnership for Advanced Computational Infrastructure. R.W. acknowledges
support from Fundaci\'on Antorchas Grants No A-13740/1-80 and A-13661/1-27.


\end{multicols}
\end{document}